\documentclass{ametsocV6.1}

\newcommand{\dd}{\mathrm{d}}
\usepackage{booktabs}

\title{Analytical insights into the transient climate response}

\authors{Boriana Chtirkova \aff{a, b} \correspondingauthor{Boriana Chtirkova, chtirkova@princeton.edu} 
}

\affiliation{\aff{a}{Department of Geosciences, Princeton University, Princeton, New Jersey}\\
\aff{b}{Program in Atmospheric and Oceanic Sciences, Princeton University, Princeton, New Jersey}
}

\abstract{The temperature in the transient climate response is lower than the equilibrium temperature for the same amount of forcing. The degree of disequilibrium is not constant in time and depends on various climate parameters. We solve the heat diffusion equation -- representing here a dynamical response to a temperature-driven potential difference rather than diffusion -- with a surface temperature feedback for linearly increasing forcing. The surface temperature initially evolves at a slower rate than the corresponding steady-state (SS) temperature, and its rate increases until reaching quasi-steady state (QSS), when the SS and QSS temperatures evolve in parallel with a constant offset. The offset depends on the rate of forcing and the total heat capacity of the system, divided by the square of the climate feedback. The timescale over which the climate system approaches QSS depends only on the effective ocean mixing and depth and is on the order of thousands of years. Over societally relevant timescales ($\sim100$ years), the top-of-atmosphere energy imbalance increases, and the actual temperature moves farther from the steady-state temperature expected for the same forcing.}

\begin{document}

\maketitle

%%%%%%%%%%%%%%%%%%%%%%%%%%%%%%%%%%%%%%%%%%%%%%%%%%%%%%%%%%%%%%%%%%%%%
% SIGNIFICANCE STATEMENT/CAPSULE SUMMARY
%%%%%%%%%%%%%%%%%%%%%%%%%%%%%%%%%%%%%%%%%%%%%%%%%%%%%%%%%%%%%%%%%%%%%
\statement
Understanding why the transient climate response differs from equilibrium climate sensitivity is essential for interpreting near-term climate change. By solving a diffusive ocean model with a surface temperature feedback under linearly increasing forcing, we show that the climate system moves progressively further from equilibrium over societally relevant timescales, accompanied by an evolving top of atmosphere energy imbalance. We provide a simple, physically transparent framework on the roles of the surface temperature feedback and effective ocean mixing in shaping the temporal evolutions of temperature and the top of the atmosphere imbalance.

%%%%%%%%%%%%%%%%%%%%%%%%%%%%%%%%%%%%%%%%%%%%%%%%%%%%%%%%%%%%%%%%%%%%%
% MAIN BODY OF PAPER
%%%%%%%%%%%%%%%%%%%%%%%%%%%%%%%%%%%%%%%%%%%%%%%%%%%%%%%%%%%%%%%%%%%%%
\section{Introduction}
The temperature response to a transient increase in the forcing is always lower than the temperature which the system will approach if the forcing is fixed after a certain point in time. To address the question whether, at the time of analysis, the observed warming is consistent with forcing, \citet{Wigley_1985} discuss an analytical solution of the heat diffusion equation that describes the evolution of the surface temperature of a system with large thermal inertia. They show that at any given point the forced climate system can be quite far from equilibrium, and they derive the degree of disequilibrium to depend on the climate feedback and the effective ocean diffusivity, representing the vertical ocean mixing. A similar problem setup has been investigated by \citet{Lebedeff_1988, Watts_1994, Morantine_1994} for either a mixed layer or a full ocean. These diffusion model studies show that the response timescale depends on the assumptions of the domain depth and the climate feedback parameter, yielding a timescale on the order of a few hundred years.

We revisit the problem of interpretation of the temperature state of the climate system using a diffusion model, but with energy conserving boundary conditions. In this setup, the heat equation becomes a continuous form of the frequently used two-box model \citep{Gregory_2000, Held_2010}. Diffusion-based energy balance models do not represent diffusion, but rather a dynamical response of the fluid driven by a potential difference -- in most cases this potential is represented by temperature. We will provide a compact analytical expression for the surface temperature and top of atmosphere imbalance response to linearly increasing forcing. The timescale we estimate for the system to approach quasi-steady state depends on the domain depth and heat exchange coefficient but not on the climate feedback parameter and is on the order of a thousand years. This long timescale explains why the evolution of surface temperature in a 150-year simulation with linearly increasing forcing appears to be linear, but its tendency is different from the steady-state temperature tendency.

\section{Background}
\subsection{The transient climate response in atmosphere-ocean general circulation models}
At present, the degree of disequilibrium is mostly inferred from atmosphere-ocean general circulation model (AOGCM) simulations. The adopted procedure is to estimate the temperature which the system approaches for a given amount of constant forcing $T_{\mathrm{ECS}}$ (e.g. from the abrupt-4xCO2 experiment) and the temperature at a fixed point of a run with monotonically increasing forcing $T_{\mathrm{TCR}}$ (e.g. from the 1pctCO2 experiment) \citep{Eyring_2016}. These two temperatures are usually normalized to a doubling of the forcing ($F_{2x}$) and are termed equilibrium climate sensitivity (ECS) and transient climate response (TCR).  Within coupled climate models, the ranges for ECS are between $1.8-5.6$ K and for TCR between $1.3-3.0$ K \citep{IPCC2023, Flynn_2020, Nijsse_2020, Meehl_2020}. The definition of $T_{\mathrm{TCR}}$ as the temperature around the year of doubling of CO$_2$ emissions for a scenario with a growth rate of 1\% per year \footnote{Concentrations grow exponentially with a rate of 1 \% per year: $C(t) = C_0 1.01^t$, the time $t'$ at which $C(t')=2C_0$ is $2=1.01^{t'}$, $t' = \log_{1.01}(2) = \frac{ln(2)}{ln(1.01)} \approx 70$.} is a practical way to assess model results. However, $T_{\mathrm{TCR}}$ is a function of the forcing growth rate and may not scale linearly with the forcing.

$T_{\mathrm{TCR}}$, unlike $T_{\mathrm{ECS}}$, is not only a function of the climate feedback parameter, but is also influenced by ocean heat uptake \citep{Hoffert_1980, Wigley_1985, Raper_2002, Gregory_2008, Rose_2016}, or more precisely, by how efficiently the ocean takes up heat over the time period used to evaluate the $T_{\mathrm{TCR}}$. The ocean heat uptake can affect the transient climate response by changing the effective heat capacity of the system \citep{Gregory_2008, Gregory_2023} and by modulating the response of the atmosphere \citep{Held_2010, Winton_2010, Rose_2016, Jeevanjee_2026}. \citet{Kostov_2014} relate the Atlantic Meridional Overturning Circulation (AMOC) with the efficiency of ocean heat uptake and show that climate models with a stronger AMOC have a larger effective heat capacity and therefore lower $T_{\mathrm{TCR}}$. \citet{Kuhlbrodt_2012} highlight the importance of the eddy-induced thickness diffusivity parameter in climate models, which is particularly effective in the Southern Ocean. The Southern Ocean is estimated to be the ocean region with the largest increase in heat content \citep{Frolicher_2015}. The efficiency with which the ocean propagates heat in depth is also related to the depth of the pycnocline \citep{Newsom_2023} and salinity distributions \citep{Liu_2023}. 

Simulations from coupled general circulation models with linearly increasing forcing indicate a linear evolution of the surface temperature in a period of 100-150 years after the onset of the forcing (e.g. \cite{Gregory_2000, Gregory_2008}). By rearranging the linear climate feedback equation: $F(t) - \lambda \Delta T(t) = N(t)$, where $\Delta T(t)$ is the departure from the unperturbed temperature, $\lambda$ is the climate feedback parameter, $F$ is the additional forcing and $N$ is the system imbalance, one obtains:
\begin{equation}
    \label{eq:tcr_basic}
    \Delta T(t)=  \frac{F(t)}{\lambda + \frac{N(t)}{T(t)}} = \frac{F(t)}{\lambda + \varkappa(t)},
\end{equation}
in which $\varkappa(t)=\frac{N(t)}{T(t)}$ is the definition of ocean heat uptake efficiency \citep{Gregory_2000, Gregory_2008, Gregory_2015, Gregory_2023}. Since both the temperature and imbalance vary in time, $\varkappa$ will depend on the period over which it is evaluated and will vary from $\varkappa \rightarrow \infty $ in the beginning of the simulation to $\varkappa \rightarrow 0$ after sufficiently long \citep{Gregory_2023}. For inter-model comparisons, it has become standard to evaluate $\varkappa$ from $T_{\mathrm{TCR}}$ around the year of doubling of the forcing and $\varkappa$ is often interpreted as an AOGCM-specific coefficient of thermal coupling between the upper and deep ocean.

\subsection{The transient climate response in a two-box diffusive model}
A conceptual framework for investigating the climate response is a two-box (or two-layer) diffusive model \citep{Gregory_2000, Held_2010, Winton_2010}, which is formulated as a system of two coupled ordinary differential equations:

\begin{equation}
\label{eq:two-box}
\left|
\begin{aligned}
    C_1\frac{dT_1}{dt} &= F - \lambda T_1 - \epsilon \gamma (T_1 - T_2) \\
    C_2\frac{dT_2}{dt} &= \gamma (T_1 - T_2)
\end{aligned}
\right.
\end{equation}

The forcing $F$ at the top of the atmosphere acts on a layer with a heat capacity $C_1$ and temperature $T_1(t)$, which transports heat to a lower layer with a heat capacity $C_2$ and temperature $T_2(t)$. The upper layer may represent the well-mixed upper ocean, and the lower layer -- the deep ocean, such that the system is commonly evaluated for $C_1 << C_2$. The heat flux between the two layers is proportional to their difference by a heat exchange coefficient $\gamma$. The system loses heat to space through a climate feedback parameter $\lambda$ acting through the upper layer temperature. The full version of this model includes an efficacy factor $\epsilon$ \citep{Held_2010, Winton_2010}, which adds a modulation to the feedback by the strength of ocean heat uptake. For the purpose of this paper we consider no such modulation, i.e. $\epsilon=1$.

 The 2-layer model (equation \ref{eq:two-box}) recovers a linear relationship between the forcing and surface temperature in the limit $C_1 \to 0$ and $T_2 \to 0$ (or $C_2 \to \infty$): 

\begin{equation}
\label{eq:limit}
    \lim_{\scriptscriptstyle C_1 \to 0, T_2 \to 0} T_1
    = \frac{F(t)}{\lambda + \gamma},
\end{equation}
but by doing this one loses all time information and there are no estimates for the validity range of this approximation \citep{Jeevanjee_2025}.

Within the 2-box model the thermal coupling coefficient $\gamma$ is a constant parameter, which unlike $\varkappa(t)$ does not vary in time; hence it is unclear what the physical interpretation of the two quantities should be. There is no standard procedure for diagnosing $\gamma$ from climate models. One way would be to suppose the limit from equation \ref{eq:limit} is valid in the year of forcing doubling and compute $\gamma$ from $T_{\mathrm{TCR}}=T_1$. A second approach is numerically fitting the coefficients of the two-layer model to AOGCM output, in which $C_1$ is often treated as a free parameter, different from zero (e.g. \citet{Geoffroy_2013}). This raises questions about how estimates of $\gamma$ obtained through different methods, i.e. supposing different heat capacities, should be interpreted.

\section{Setup and solutions of the heat equation}
\label{sec:math}
We set up the classical heat (diffusion) equation in which temperature is a function of time and depth $\mathcal{T} = \mathcal{T}(t, z)$:

\begin{equation}
\label{eq:diffusion}
c \frac{\partial  \mathcal{T}}{\partial t} = K \frac{\partial ^2 \mathcal{T}}{\partial z^2}
\end{equation}

A significant part of the analysis to follow will focus on the surface temperature, denoted $T(t) = \mathcal{T}(t, z=0)$. For seawater, the volumetric heat capacity is $c \approx 4 \times 10^6$ Jm$^{-3}$K$^{-1}$  and the molecular heat conductivity is $K \approx 0.6$ WK$^{-1}$m$^{-1}$ \citep{Alexandrov_2017}. The molecular diffusivity is negligible compared to the effects of advection, convection, and mixing arising from wind stress, eddies, tides and bottom boundary layers. This effective mixing can be expressed as effective ocean diffusivity $\kappa$ [m$^2$s$^{-1}$] \citep{Redi_1982, Gent_1990, Gnanadesikan_1999, Marshall_2014}. 
Estimates of global ocean diffusivity are usually built on theories which balance buoyancy with eddy processes and deep water formation. \cite{Gnanadesikan_1999} predict the mean ocean stratification by linking the low-latitude upwelling to Northern Hemisphere sinking and Southern ocean eddies. \cite{Marshall_2014} link the processes which control this mean stratification to ocean heat uptake under climate change. 
Theoretical and observational studies suggest a vertical structure of $\kappa$ being $\sim 10^{-5}$ m$^2$s$^{-1}$ at mid-depths and $\sim 10^{-4}$ m$^2$s$^{-1}$ in the abyss \citep{Munk1966, Munk1998, Nikurashin_2011, Rogers_2023}. A representative value of $\kappa$ for the upper ocean is $2.5 \times 10^{-5}$ m$^2$s$^{-1}$ and below 2500 meters it becomes larger than $2 \times 10^{-4}$ m$^2$s$^{-1}$ due to the influence of orography. Relating these estimates of diffusivity directly to temperature (i.e. $\kappa = \frac{K}{c}$) yields effective heat conductivity $K \approx 100-800$ WK$^{-1}$ m$^{-1}$. This coefficient describes the dynamical response of the ocean to a potential difference and is not to be interpreted as molecular diffusivity.

The special case $K(z)=K$ allows one to find an analytical solution that reveals the scalings needed to guide interpretation of climate model simulations where, as in the real ocean, $K$ may be a function of depth, geographical location, and time.

For an energy conserving system, the boundary conditions need to be formulated in terms of the fluxes at the top and bottom. The upper boundary is formulated using the forcing and a linear feedback so it contains both the function and its derivative in $z$ (Robin boundary condition). The lower boundary condition says there is no flux out of the lower boundary (Neumann boundary condition). 
The zero-flux lower boundary condition ensures energy conservation and differs from the $T=0$ lower boundary condition used in the preceding literature \citep{Wigley_1985, Lebedeff_1988, Morantine_1994, Watts_1994}, the implications of which are further discussed with the solution. The boundary conditions are formulated as:

\begin{equation}
\label{eq:bcs}
\begin{aligned}
  z = 0:\quad \mathcal{F}(t,0) &= -K \left.\frac{\partial\mathcal{T}}{\partial z}\right|_{z=0} \\
                               &= F(t) - \lambda\,\mathcal{T}(t,0) = N(t) \\[4pt]
  z = L:\quad \mathcal{F}(t,L) &= -K \left.\frac{\partial\mathcal{T}}{\partial z}\right|_{z=L} = 0,
\end{aligned}
\end{equation}

where $\lambda > 0$ is the climate feedback parameter. In this formulation, the upper boundary condition sets the heat tendency of the entire system. In climate science, it is referred to as the net top of the atmosphere (TOA) radiative flux, and for $N(t) = F(t) - \lambda T(t) \neq 0$ it is the TOA energy imbalance \footnote{The correct expression is for the temperature difference $\Delta T$ relative to the pre-industrial global mean surface temperature: $N(t) = F(t) - \lambda \Delta T(s)$. For simplicity, we set the initial temperature to 0 and omit the $\Delta$ throughout the paper.}. The values of the climate feedback parameter estimated from climate models are in the range $0.6-1.5$ Wm$^{-2}$K$^{-1}$ \citep{IPCC2023, Flynn_2020, Nijsse_2020}. An appropriate value for the domain depth would be the effective ocean depth~\footnote{From the mean depth of the ocean 3682 m times its surface area fraction 71\%.} $L \approx 2600$ m. We adopt the convention of the flux being positive downward and $z$ to increase in depth ($z>0$), hence the minus signs in the boundary conditions (equation \ref{eq:bcs}).

\subsection{Steady-state solution}

We first consider the  case with constant forcing $F(t>0) = const$. After sufficiently long the system will approach a steady state (SS), defined by $\frac{\partial \mathcal{T}}{\partial t}=0$, in which the fluxes are in equilibrium and the temperature is constant in depth: 
\begin{equation}
    \label{eq:ss_solution}
    \mathcal{T}_\mathrm{SS} (t, z) = T_{\mathrm{SS}} = \frac{F}{\lambda}
\end{equation}
For $F=F_{2x}$, $T_\mathrm{SS}$ corresponds to the equilibrium climate sensitivity $T_{\mathrm{ECS}}$ commonly used in climate science. 

We can extend the definition of steady-state temperature for cases in which $F=F(t)$. Then, $T_{\mathrm{SS}}(t) = \frac{F(t)}{\lambda}$ will be the temperature the system will eventually approach if the forcing is fixed at its value at time $t$.

For other forcing cases, such as a sudden return to preindustrial conditions or a linear forcing ramp for a finite duration followed by steady forcing, the system will eventually approach flux equilibrium with the same solution (equation \ref{eq:ss_solution}), in which $F$ is the final forcing.

\subsection{Quasi-steady state solution}

For a linearly increasing forcing $F = at$, $\frac{dF}{dt} = a$, the system 
possesses a solution, in which the temperature tendency across the whole domain is constant $\frac{\partial \mathcal{T}}{\partial t} = const = \frac{a}{\lambda}$: a quasi-steady state (QSS). The QSS temperature is:

\begin{equation}
\label{eq:qss}
\mathcal{T}_\mathrm{QSS}(t, z) = \frac{-2 a c K L + 2 a \lambda K t - 2 a \lambda c L z + a \lambda c z^2}{2 \lambda^2 K} = \frac{a}{\lambda}t + f(z)
\end{equation}

with $f(z) = \frac{-2 a C K L - 2 a \lambda c L z + a \lambda c z^2}{2 \lambda^2 K}$. The terms containing $t$ and $z$ are separated: the temperature tendency is $\frac{\partial \mathcal{T}(t, z)}{\partial t} = \frac{a}{\lambda}$ and the corresponding function of depth $\mathcal{T}(z)=f(z)$ is quadratic. 

We can express the surface temperature and its offset from $T_\mathrm{SS}$ for each amount of forcing $T_\mathrm{SS}(t) = \frac{a}{\lambda}t$ as:

\begin{equation}
    \mathcal{T}_\mathrm{QSS}(t, z=0) = T_\mathrm{QSS}(t) = \frac{a}{\lambda}t - \frac{acL}{\lambda^2} = T_\mathrm{SS}(t) - \frac{acL}{\lambda^2} 
\end{equation}

We note that at the surface $T_\mathrm{QSS}(t)$ is independent of heat conductivity $K$, but the time scale to reach the quasi-steady state is a function of $K$ (see next section). This is different from solutions to the diffusion equation with a lower boundary condition $\mathcal{T}(t, z \rightarrow \infty)=0$ as in \citet{Wigley_1985}, where such a QSS is never reached. The lower boundary condition $\mathcal{T}(t, z \rightarrow L)=0$ used in \citet{Lebedeff_1988, Morantine_1994} allows the system to reach a QSS but this QSS additionally depends on the heat conductivity coefficient $K$ and its corresponding vertical structure function $\mathcal{T}(z)$ includes $z^3$ terms (in addition to the $z^2$ terms in our solution). The same surface temperature QSS without $K$ is derived as an asymptote for the diffusion equation solution by \citet{Watts_1994}. In the next section, we suggest a time scale to approach this QSS that is different from the one in \citet{Watts_1994}.

\begin{figure}[h!]
    \centering
    \includegraphics[width=\linewidth]{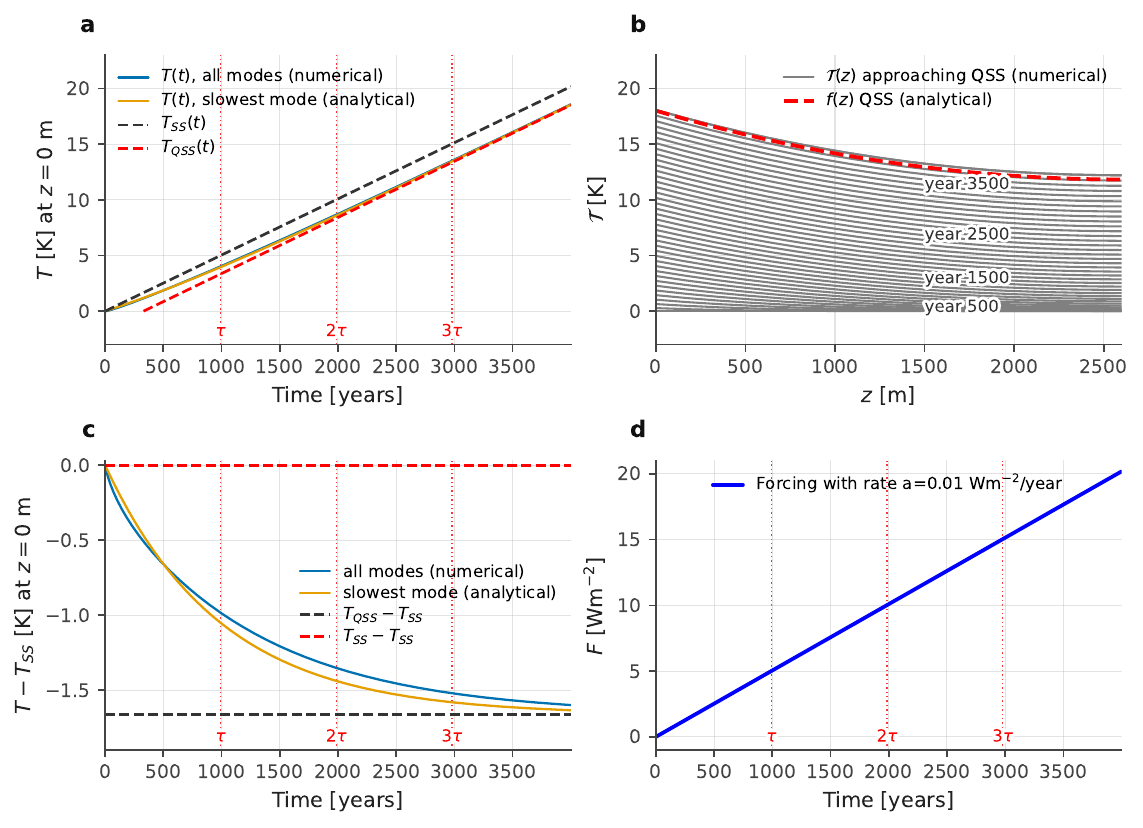}
    \caption{Analytical (slowest decaying mode only) and numerical (all decaying modes) solutions to the heat equation with parameters $F_{2\times}=3.5$ Wm$^{-2}$, $\lambda=1$ Wm$^{-2}$K$^{-1}$, $K=350$\,Wm$^{-1}$K$^{-1}$, $L=2600$\,m and a forcing increase of 1\,permille/year. Evolution of surface temperature anomaly (a), corresponding temperature profiles in depth (b) - individual gray lines are 100\,years apart, evolution of surface temperature anomaly relative to the steady state (SS) temperature (c), evolution of the forcing (d).}
    \label{fig:ts}
\end{figure}

Figure\,\ref{fig:ts}\,a,c presents $T_\mathrm{SS}(t)$ and its parallel $T_\mathrm{QSS}(t)$ at the surface (i.e. $z=0$) as black and red dashed lines and Figure\,\ref{fig:ts}\,b -- the quadratic temperature profile in depth $f(z)$ as red dashed line.

Since the temperature tendency is constant, the corresponding imbalance is also constant in time:

\begin{equation}
    N_\mathrm{QSS} = -K \left.\frac{\partial \mathcal{T}}{\partial z}\right|_{z=0} = -K \left.\frac{-2a \lambda c L + 2 a \lambda c z}{2\lambda^2K}\right|_{z=0} = \frac{acL}{\lambda}
\end{equation}

$N_\mathrm{QSS}$ is time independent and depends on the system heat capacity and the rate of forcing. For an equivalent ocean depth $L=2600$ m and approximately the current rate of greenhouse gas forcing $a=0.04$ Wm$^{-2}$yr$^{-1}$ \citep{cmip6_forcing}, the TOA imbalance at QSS is $N_\mathrm{QSS} \approx \frac{13}{\lambda}$\,Wm$^{-2}$, which for any likely value of $\lambda$ is higher than the currently observed $1.0 \pm 0.2$\,Wm$^{-2}$ \citep{Loeb_2021, Loeb_2024}, indicating that even after $\sim 100$ years of approximately linearly increasing forcing, Earth has not approached QSS. In the following, we show that this result is consistent with the analytical solution for the timescale to reach QSS in the diffusion model (equations \ref{eq:diffusion}, \ref{eq:bcs}).

\subsection{Approach to the QSS solution}
If we add a zero initial condition to the system $\mathcal{T}(t=0, z) = 0$, the full solution is an infinite sum of time-decaying exponential modes:

\begin{equation}
    \mathcal{T}(t, z) = \frac{a}{\lambda}t - \sum_{i=1}^{\infty} f_i(z) e^{-t/\tau_i},
\end{equation}

where $\tau_i$ and $f_i(z)$ are the individual time-decaying modes and their corresponding vertical structure functions.

In Appendix~A we show that in the limit $\lambda L >> K$, one can approximate the timescale of the slowest decaying mode $\tau_{sl}$  as
\begin{equation}
\label{eq:tau}
\tau_{sl} \approx \frac{c}{K}\frac{4L^2}{\pi^2}
\end{equation}

The timescale $\tau_{sl}$ to approach the QSS with the quadratic form of the vertical structure function $f(z)$ is a property of the system (i.e. a function of $c, K, L$), and is independent of the forcing rate $a$ or total forcing $at$. Inserting numerical values ($L=2600$m, $K=350$Wm$^{-1}$K$^{-1}$, $c=4\times10^6$Jm$^{-3}K^{-1}$), we obtain $\tau \approx 990~\mathrm{years}$.

The approximation of the full solution with one decaying mode thus yields for the evolution of temperature and TOA energy imbalance:
\begin{align}
\label{eq:qss_approach}
    \mathcal{T}(t, z) &\approx \frac{a}{\lambda}t - (1-e^{-t/\tau})f(z) \\
    \label{eq:qss_approachn}
    N(t) &\approx (1-e^{-t/\tau})\frac{acL}{\lambda}.
\end{align}

With the slowest mode approximation, the decaying mode $\tau = \tau_{sl}$ is the slowest decaying mode of the diffusive system and the term $cL$ is the total heat capacity of the system $C$. The general form of the solutions \ref{eq:qss_approach}-\ref{eq:qss_approachn} is valid for not only the diffusion equation. The one-layer model $C \frac{dT}{dt} = F(t) - \lambda T(t)$ recovers the same solution but with a response timescale: $\tau = C/\lambda$ (see also \citet{Roe_2009}). The two-box model (equations \ref{eq:two-box}) also has the same solution form, with the repose timescale $\tau$ depending on the two heat capacities, the heat exchange coefficient and the climate feedback parameter (see, for example, Appendix B.3 in \citet{Gregory_2023}). The timescale provided by \citet{Watts_1994} also depends on both $C$ and $\lambda$. The disappearance of $\lambda$ in the response timescale here is due to the approximation $\lambda L >> K$. The timescale in a model with a diffusive domain is always longer than the timescale in the simplest one-layer model.

\section{The growing apparent heat capacity}
\label{sec:intuition}

One can develop intuition for the transient response to monotonically increasing forcing by thinking of the system as a single layer with temperature $T$, which is also the surface temperature, and a time-evolving heat capacity:

\begin{equation}
    C_\mathrm{app}(t) \frac{\dd T}{\dd t} = N(t) = F(t) - \lambda T(t)
\end{equation}

Here, the apparent heat capacity $C_\mathrm{app}$ is not a real physical heat capacity but arises as a mathematical concept describing the time-dependent influence of the diffusive domain. The notion of a time-varying heat capacity has been shown to better capture the climate response in AOGCMs as compared to a single, constant in time, heat capacity (e.g. \citet{Watterson_2000, Donohoe_2014}). In our case, we simply diagnose $C_{\mathrm{app}}$ as: $C_\mathrm{app} = \frac{N}{\dd T/\dd t} = \frac{N \dd t}{\dd T} = \frac{E}{\Delta T}$, where $E$ is the accumulated energy in the system in Joules. The apparent heat capacity may be diagnosed from the instantaneous flux and temperature tendency, $C_\mathrm{app} = \frac{N}{\dd T/\dd t}$, or from the integrated change in heat content $E$ and the surface temperature change $\Delta T_s$, $C_\mathrm{app} = \frac{\Delta E}{\Delta T}$. Equivalently, we may define an apparent depth $L_\mathrm{app} = C_\mathrm{app}/c$, which is the apparent heat capacity expressed as equivalent ocean depth in meters.

\begin{figure}[h!]
    \centering
    \includegraphics[width=\linewidth]{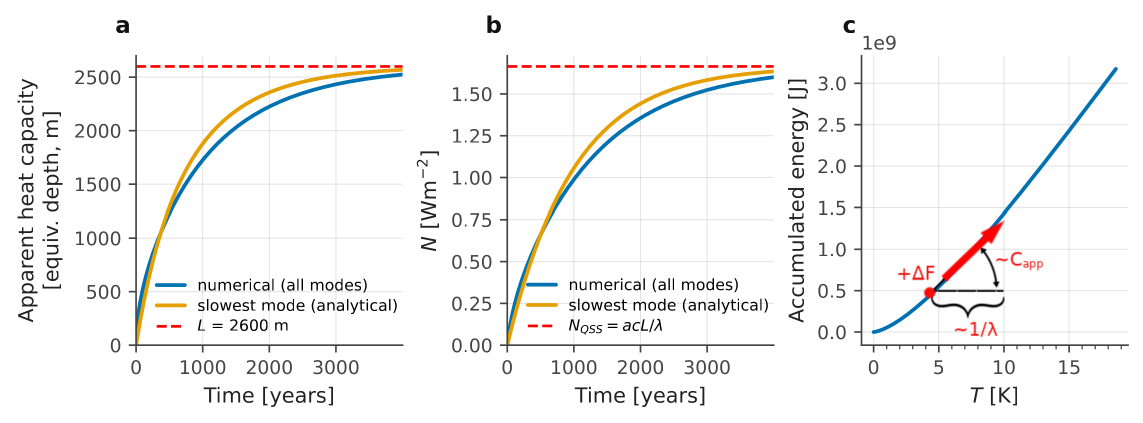}
    \caption{Temporal evolution of the apparent heat capacity (a) and top of domain imbalance (b). Phase space of accumulated energy and surface temperature (c). Parameters are the same as in Figure \ref{fig:ts}.}
    \label{fig:imb}
\end{figure}

Using the expressions for $T(t)=\mathcal{T}(t, z=0)$ and $N(t)$ from equations \ref{eq:qss_approach} and \ref{eq:qss_approachn}, we can derive $L_\mathrm{app}(t)$ for the case of linearly increasing forcing:
\begin{equation}
\label{eq:capp}
L_\mathrm{app} = \frac{\lambda \left(-1 + e^{t/\tau}\right) L \tau}{-c L + \lambda \tau\, e^{t/\tau}} \xrightarrow[t \to \infty]{} L,
\end{equation}

in which the dependence of the exact forcing rate $a$ disappears. The increase of $L_\mathrm{app}$ over time indicates that close to the onset of the forcing growth, the system responds as a shallow reservoir and over time approaches the limit of the total domain depth $L$. This happens in tandem with the growing imbalance $N$. In QSS, $L_\mathrm{app} = L$ and $N=\frac{acL}{\lambda}$, and prior to QSS, both $L_\mathrm{app}$ and $N$ increase logarithmically (see Figure \ref{fig:imb} a, b).

A simple way to think of the TCR problem is as partitioning a $\Delta F$ over a short period of time ($\Delta t \sim 1$ year) in E($T$) phase space, which pictures the total accumulated energy $E$ versus the change in surface temperature $T$ (Figure \ref{fig:imb} c). A part of the additional amount of forcing $\Delta F$ will raise $N$, i.e. introduce a temperature tendency in the deeper layers and result in energy accumulation in the system, and the other part raises the surface temperature $T$ and through the climate feedback decrease the imbalance $N$. The angle in the $E(T)$ phase space is $C_\mathrm{app}(t)$ and the projection on the $T$-axis is proportional to $1/\lambda$. Before QSS, $C_\mathrm{app}(t)$ depends not only on $L$ but also on $K$ and $\lambda$ (equation \ref{eq:capp}).

It is important to mention that $C_\mathrm{app}$ depends on the functional form of the forcing. For a Heaviside function ($F(t\leq 0) = 0, F(t>0)=\mathrm{const}$ (e.g. the abrupt-4xCO2 experiment), $L_\mathrm{app}$ will evolve initially from a very small value to a maximum that exceeds $L$ as the deeper layers still experience a substantial temperature tendency, but the surface temperature is close to the steady state temperature. This limit is further discussed in Appendix~C.

\section{Application for interpreting models and observations}
\label{sec:models}

\subsection{The ratio between the surface temperature tendency and $\frac{\dd T_{\mathrm{SS}}}{ \dd t}$ in AOGCMs}

We now use the analytical insights from the previous sections to interpret the temporal evolution of surface temperature and top of atmosphere imbalance in the 1pctCO2 experiment. We wish to characterize the degree to which the system has approached QSS, which depends on the time at which the temperature tendency is evaluated ($t'$) relative to the response time scale of the system ($\tau$). A convenient metric is the ratio of the surface temperature tendency ($\frac{\dd T}{\dd t}(t)$), which is time-dependent, relative to the corresponding steady state temperature tendency ($\frac{\dd T_\mathrm{SS}}{\dd t}$). The latter can be simply estimated using $F_{2x}$ and $\lambda$ from an abrupt-4xCO2 run: one can derive the forcing rate from $F_{2x}$ as $F(t) = F_{2x} \frac{ln(1.01^t)}{ln(2)}, a = \frac{\dd F}{\dd t} = F_{2x} \frac{ln(1.01)}{ln(2)}$ as $\frac{\dd T_\mathrm{SS}}{\dd t}=\frac{a}{\lambda}$. The equilibration time scale of the system is much longer than the 150-year length of 1pctCO2 simulations. It is thus possible to approximate $\frac{\dd T}{\dd t}(t)$ to the linear regression of the surface temperature over the entire 150-year period or any subset large enough to not be influenced by internal variability. It is not possible to obtain sufficient information to estimate the total heat capacity of the system from AOGCM simulations and we need to make an approximation for the total domain depth $L$: we set $L=2600$m as the depth of the ocean column if all water on Earth was evenly distributed over a flat surface, then the total heat capacity of the system $C=cL$ will correspond to the total heat capacity of Earth's oceans. 

\begin{figure}[h!]
    \centering
    \includegraphics[width=\linewidth]{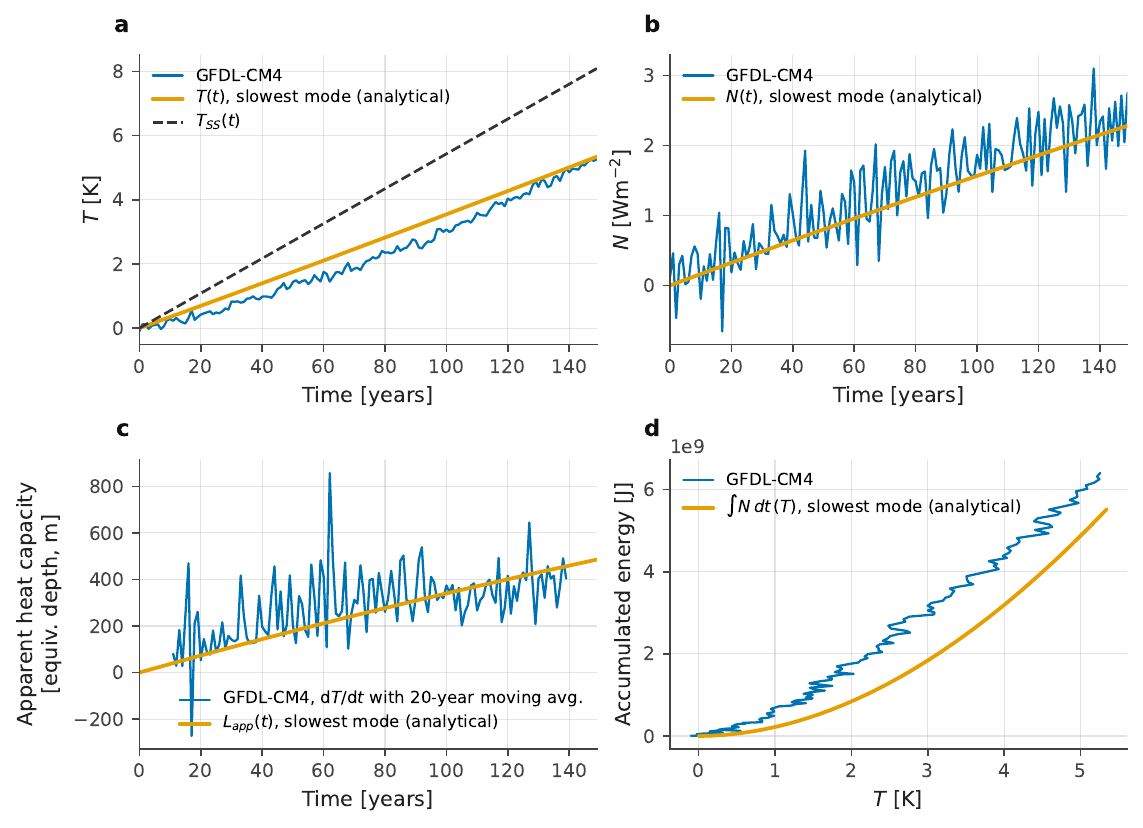}
    \caption{Temporal evolution of surface temperature (a), top of atmosphere imbalance (b), apparent heat capacity as equivalent depth (c) and accumulated energy and surface temperature phase space (d). Blue lines correspond to the GFDL-CM4 1pctCO2 run (anomalies relative to piControl). Orange lines are the analytical approximation (equations \ref{eq:qss_approach}-\ref{eq:qss_approachn}) with $F_{2x}=3.13$ Wm$^{-2}$, $\lambda=0.83$ Wm$^{-2}$K$^{-1}$, $K=318$ Wm$^{-1}$K$^{-1}$, $L=2600$ m.}
    \label{fig:cm4}
\end{figure}

Figure \ref{fig:cm4} a shows the surface temperature under a 1pctCO2 scenario -- the linear regression will not capture the slight curvature in this evolution, but it is a good indicator for the state of the system as the curvature is due to faster decaying modes and our analytical approximation is only derived for the slowest one.

We use equation \ref{eq:qss_approach} to estimate the ratio of the derivatives analytically at any point $t'$:

\begin{equation}
\label{eq:for_k}
\frac{\dd T}{\dd t}(t=t')\Big/\frac{\dd T_\mathrm{SS}}{\dd t} = 1 - \frac{K e^{-t'/\tau}\pi^2}{4 \lambda L},
\end{equation}

which is the time-dependent correction to apply to $\dd T_\mathrm{SS}/\dd t$ in order to estimate the transient tendency $\frac{\dd T}{\dd t}(t=t')$ at time $t'$. 

This equation shows how much the surface temperature tendency is reduced in the transient response. \citet{Roe_2009} has shown that models with higher climate sensitivity (smaller $\lambda$) have a more suppressed response as compared to their $\frac{\dd T_{\mathrm{SS}}}{ \dd t}$. We show that this suppression is not due to their response time scale, which we find to be independent of $\lambda$ (section \ref{sec:math}), but to the offset between $T(t)$ and $T_{\mathrm{QSS}}$ resulting in the proportionality $\frac{\dd T}{\dd t}(t=t')/\frac{\dd T_\mathrm{SS}}{\dd t} \sim \frac{1}{\lambda}$ for $t'<\tau$.

The relation in equation \ref{eq:for_k} can be used in two directions. Forward, assuming $K=350$Wm$^{-1}$K$^{-1}$, the ratio evolves from $1 - \frac{K\pi^2}{4 \lambda L} \approx 0.67$ near the onset of the forcing to 1 as system approaches QSS. In reverse, we can instead diagnose $K$ from a simulation (see the following section \ref{sec:K_from_cmip6}): taking the ratio $\frac{\dd T}{\dd t}(t=t') \big/ \frac{\dd T_\mathrm{SS}}{\dd t}$ at time $t'$ near the forcing onset in the 1pctCO2 run to be 0.67 and solving for $K$ yields $K=350$Wm$^{-1}$K$^{-1}$. This falls within the wide theoretical range of $100-800$Wm$^{-1}$K$^{-1}$ (see section \ref{sec:math}), though the specific value of $350$ should be understood as diagnosed from AOGCMs rather than theoretically expected.

Going one step further, one can use equations \ref{eq:qss_approachn} and \ref{eq:for_k} to obtain the expression:
\begin{equation}
\label{eq:rates}
    \frac{dT/dt}{dT_\mathrm{SS}/dt} = 1 - \frac{dN/dt}{dF/dt},
\end{equation}

which relates the rate of change of temperature to the rate of change of the TOA imbalance in the transient run. The ratios of these derivatives are obtained as linear regressions from models within the Coupled Model Intercomparison -- Phase 6 (CMIP6, \citet{Eyring_2016}) and shown in the last two columns of Table \ref{tab}. On average, the transient temperature evolves at 67\% the rate of the SS temperature, and the ratio between the rates of change of $N$ and $F$ is therefore 33\%. Over time, $\frac{\dd T/\dd t}{\dd T_\mathrm{SS}/\dd t}$ will increase at the expense of $\frac{\dd N/\dd t}{\dd F/\dd t}$.

If the ocean heat uptake efficacy ($\epsilon$ in equation \ref{eq:two-box}) is substantially different from 1, one needs to account for it in equation \ref{eq:rates} as (see equation 15 in \citet{Winton_2010}):

\begin{equation}
    \frac{\dd T/\dd t}{\dd T_\mathrm{SS}/\dd t} = 1 - \epsilon \frac{\dd N/\dd t}{\dd F/\dd t},
\end{equation}

Therefore, we can expect that models, for which the last two columns ($\frac{\dd T}{\dd T_\mathrm{SS}} + \frac{\dd N}{\dd F}$) in Table \ref{tab} do not  sum to 1, to have $\epsilon \neq 1$.  

% \usepackage{booktabs}
\begin{table}
\caption{Heat equation parameters estimated from the 1pctCO2 and abrupt-4xCO2 runs by coupled models. The first 3 columns are copied from \cite{Flynn_2020}. $K$ is estimated from eq. \ref{eq:for_k} and $\gamma$ from eq. \ref{eq:gamma}. The estimates of $K$ and $\gamma$ depend on the assumption for total domain depth $L=2600$m. To estimate $\gamma$, we also suppose a ratio between the upper ($C_1$) and deep layer ($C_2$) heat capacities $k=C_1/C_2=0.1$. $dT$ is the rate of temperature change, $dT_{SS}$ is the rate of the corresponding temperature at flux equilibrium, $dN$ is the rate of change of TOA imbalance, $dF$ is the rate of forcing. The ratios of derivatives are based on linear regressions computed over the full length of the 1pctCO2 run (150 years).}
\label{tab}
\tiny
\begin{tabular}{@{}lrrrrrrr@{}}
\toprule
                                                                                           & \multicolumn{1}{c}{\begin{tabular}[c]{@{}c@{}}TCR\\ {[}K{]}\end{tabular}} & \multicolumn{1}{c}{\begin{tabular}[c]{@{}c@{}}$F_{2x}$\\  {[}Wm$^{-2}$K$^{-1}$]\end{tabular}} & \multicolumn{1}{c}{\begin{tabular}[c]{@{}c@{}}$\lambda$\\  {[}Wm$^{-2}$K$^{-1}$]\end{tabular}} & \multicolumn{1}{c}{\begin{tabular}[c]{@{}c@{}}K\\ {[}Wm$^{-1}$K$^{-1}$]\\  (L=2600)\end{tabular}} & \multicolumn{1}{c}{\begin{tabular}[c]{@{}c@{}}$\gamma_{QSS}$\\ {[}Wm$^{-2}$K$^{-1}$]\\ (L=2600, k=0.1)\end{tabular}} & \multicolumn{1}{c}{dT/dT$_{SS}$} & \multicolumn{1}{c}{dN/dF} \\ \midrule
BCC-ESM1                                                                                   & 1.77                                                                      & 3.02                                                                                            & 0.92                                                                                             & 323                                                                                                 & 0.35                                                                                                                    & 0.69                         & 0.33                      \\
BCC-CSM2-MR                                                                                & 1.60                                                                      & 3.06                                                                                            & 1.00                                                                                             & 326                                                                                                 & 0.36                                                                                                                    & 0.71                         & 0.28                      \\
CESM2                                                                                      & 1.99                                                                      & 3.19                                                                                            & 0.62                                                                                             & 332                                                                                                 & 0.36                                                                                                                    & 0.53                         & 0.42                      \\
CESM2-WACCM                                                                                & 1.92                                                                      & 3.26                                                                                            & 0.62                                                                                             & 365                                                                                                 & 0.40                                                                                                                    & 0.48                         & 0.43                      \\
CNRM-ESM2-1                                                                                & 1.82                                                                      & 2.96                                                                                            & 0.62                                                                                             & 304                                                                                                 & 0.33                                                                                                                    & 0.56                         & 0.45                      \\
CNRM-CM6-1                                                                                 & 2.23                                                                      & 3.70                                                                                            & 0.77                                                                                             & 343                                                                                                 & 0.38                                                                                                                    & 0.61                         & 0.39                      \\
CNRM-ESM2-1                                                                                & 2.75                                                                      & 3.68                                                                                            & 0.66                                                                                             & 284                                                                                                 & 0.31                                                                                                                    & 0.62                         & 0.39                      \\
CNRM-CM6-1                                                                                 & 2.91                                                                      & 3.28                                                                                            & 0.62                                                                                             & 189                                                                                                 & 0.21                                                                                                                    & 0.72                         & 0.30                      \\
EC-Earth3-Veg                                                                              & 2.76                                                                      & 3.34                                                                                            & 0.80                                                                                             & 240                                                                                                 & 0.26                                                                                                                    & 0.73                         & 0.26                      \\
GFDL-CM4                                                                                   & -                                                                         & 3.14                                                                                            & 0.83                                                                                             & 318                                                                                                 & 0.35                                                                                                                    & 0.66                         & 0.35                      \\
GFDL-ESM4                                                                                  & -                                                                         & 3.84                                                                                            & 1.50                                                                                             & 408                                                                                                 & 0.45                                                                                                                    & 0.76                         & 0.25                      \\
GISS-E2-1-G                                                                                & 1.66                                                                      & 3.84                                                                                            & 1.48                                                                                             & -                                                                                                   & -                                                                                                                       & -                            & -                         \\
GISS-E2-1-H                                                                                & 1.81                                                                      & 3.47                                                                                            & 1.16                                                                                             & 351                                                                                                 & 0.38                                                                                                                    & 0.73                         & 0.29                      \\
HadGEM3-GC31-LL                                                                            & 2.47                                                                      & 3.48                                                                                            & 0.64                                                                                             & 263                                                                                                 & 0.29                                                                                                                    & 0.63                         & 0.39                      \\
INM-CM4-8                                                                                  & 1.30                                                                      & 2.64                                                                                            & 1.46                                                                                             & 204                                                                                                 & 0.22                                                                                                                    & 0.87                         & 0.13                      \\
IPSL-CM6A-LR                                                                               & 2.39                                                                      & 3.39                                                                                            & 0.75                                                                                             & 301                                                                                                 & 0.33                                                                                                                    & 0.64                         & 0.32                      \\
MIROC-ES2L                                                                                 & 1.51                                                                      & 4.03                                                                                            & 1.51                                                                                             & 478                                                                                                 & 0.52                                                                                                                    & 0.73                         & 0.30                      \\
MIROC6                                                                                     & 1.58                                                                      & 3.61                                                                                            & 1.39                                                                                             & 469                                                                                                 & 0.51                                                                                                                    & 0.71                         & 0.31                      \\
MPI-ESM1-2-HR                                                                              & 1.57                                                                      & 3.60                                                                                            & 1.27                                                                                             & 395                                                                                                 & 0.43                                                                                                                    & 0.73                         & 0.29                      \\
MRI-ESM2-0                                                                                 & -                                                                         & 3.37                                                                                            & 1.08                                                                                             & 491                                                                                                 & 0.54                                                                                                                    & 0.61                         & 0.36                      \\
NESM3                                                                                      & -                                                                         & 3.78                                                                                            & 0.84                                                                                             & 314                                                                                                 & 0.34                                                                                                                    & 0.67                         & 0.33                      \\
NorCPM1                                                                                    & 1.55                                                                      & 3.58                                                                                            & 1.29                                                                                             & 425                                                                                                 & 0.46                                                                                                                    & 0.71                         & 0.31                      \\
NorESM2-LM                                                                                 & 1.48                                                                      & 3.44                                                                                            & 1.38                                                                                             & 538                                                                                                 & 0.59                                                                                                                    & 0.67                         & 0.29                      \\
SAM0-UNICON                                                                                & 2.08                                                                      & 3.85                                                                                            & 1.05                                                                                             & 401                                                                                                 & 0.44                                                                                                                    & 0.67                         & 0.33                      \\
UKESM1-0-LL                                                                                & 2.79                                                                      & 3.56                                                                                            & 0.67                                                                                             & 286                                                                                                 & 0.31                                                                                                                    & 0.62                         & 0.38                      \\ \midrule
Mean                                                                                       & 1.98                                                                      & 3.44                                                                                            & 1.00                                                                                             & 348                                                                                                 & 0.38                                                                                                                    & 0.67                         & 0.33                      \\
5\textsuperscript{th} to 95\textsuperscript{th} p & {[}1.5, 2.8{]}                                                            & {[}3, 3.8{]}                                                                                    & {[}0.6, 1.5{]}                                                                                   & {[}210, 489{]}                                                                                      & {[}0.2, 0.5{]}                                                                                                          & {[}0.5, 0.8{]}               & {[}0.3, 0.4{]}            \\ \bottomrule
\end{tabular}
\end{table}

\subsection{Estimating the heat conductivity coefficient $K$ from $\frac{\dd T}{\dd t}(t) / \frac{\dd T_\mathrm{SS}}{\dd t}$ in AOGCMs}
\label{sec:K_from_cmip6}
An equivalent approach to evaluating the time-dependent ratio $\frac{\dd T}{\dd t}(t) / \frac{\dd T_\mathrm{SS}}{\dd t}$ would be to solve equation \ref{eq:for_k} for K. We choose $t'$ from the middle of the period over which we estimate the trend (we do this over the whole length of the 1pctCO2 run: 150 years, therefore $t'=75$years). Knowing $a$ and $\lambda$ (we take the values estimated by \citet{Flynn_2020}) and supposing that $c=4 \times 10^6$ Jm$^{-3}$K$^{-1}$ and $L=2600$m, we can solve equation \ref{eq:for_k} for K. K also appears in $\tau$ and we solve the transcendental equation, but since $\tau$ is very large, a similar solution would be obtained with $t'=0$.

The results for individual models are presented in Table \ref{tab} and the average value of K obtained from CMIP6 models is around $350$~Wm$^{-1}$K$^{-1}$.

If we return to the assumptions of $K$ being constant in depth and of analytically handling the system with the slowest decay mode instead of a continuum, the values we obtain for $K$ are likely overestimates. This is because in the initial 150-year period, the surface temperature evolution would be affected by the wind-driven mixing in the upper ocean layers, which likely corresponds to a larger $K$ than the one at mid-depths. Similarly, retaining the full continuum of modes rather than only the slowest one affects the curvature during the early period: the surface temperature initially warms more slowly under the full continuum than it would with only the slowest mode retained (see Figure \ref{fig:ts} c); this is valid for a continuum derived with constant coefficients in depth $K(z)=K$). 

To gain a quantitative sense of this estimate, it will be of use to relate $K=350$~Wm$^{-1}$K$^{-1}$ to the exchange coefficient $\gamma$ between the two layers in the two-box model (equation \ref{eq:two-box}). This can be done because the two-box model QSS can be analytically derived from the QSS solution of the heat equation (equation \ref{eq:qss}) by defining $T_1(t)$ and $T_2(t)$ and as the integrals of $T_\mathrm{QSS}(t, z)$ over the appropriate depth intervals. The full derivation is provided in Appendix B. The estimated $\gamma_{\mathrm{QSS}}$ would be a consequence of the assumptions of domain depth ($C_1 + C_2$) and ratio of $k=C_1/C_2$ and will evolve in time. We assume a domain depth of 2600 m and a ratio of 0.1. The average value of $\gamma_\mathrm{QSS}$ that we obtain in this way is around $0.4$ Wm$^{-2}$K$^{-1}$.

\citet{Geoffroy_2013} have fitted the surface temperature evolution in the abrupt-4xCO2 experiment to the 2-box model. The estimates of $\gamma$ obtained by them within CMIP5 models are in the range $0.5-1.1$ Wm$^{-2}$K$^{-1}$. Our estimate of $\gamma_\mathrm{QSS}=0.4$ Wm$^{-2}$K$^{-1}$ cannot be directly compared to the one from \citet{Geoffroy_2013} for two reasons: (1), we estimate $\gamma_\mathrm{QSS}$ which is the value $\gamma$ would approach over time. Figure \ref{fig:2box} indicates that when the heat equation is reduced to two boxes, initially $\gamma$ is larger and decreases to $\gamma_\mathrm{QSS}$; (2), we do not have a good estimate for the true depth of the domain -- we suppose $L=2600$ m based on the equivalent ocean depth. The domain depth estimated by \citet{Geoffroy_2013} is on the order of 700 m but this might be biased low due to the time-dependent effective heat capacity of the system and the finite length of the simulation  - see Figure \ref{fig:cm4} c. Indeed, if we fit the first 150 years of the numerical solution of the diffusion model (equations \ref{eq:diffusion}, \ref{eq:bcs}) with parameters $\lambda=1$~Wm$^{-2}$K$^{-1}$, $K=350$\,Wm$^{-1}$K$^{-1}$, $L=2600$\,m to the two-box model (equations \ref{eq:two-box}), we obtain the parameter values $\lambda=1$~Wm$^{-2}$K$^{-1}$, $\epsilon \approx 1.3$~Wm$^{-2}$K$^{-1}$, $C_1 \approx 100$m, $C_2 \approx 550$m, $\gamma \approx 0.8$Wm$^{-2}$K$^{-1}$, which are consistent with the values obtained in \citet{Geoffroy_2013, Geoffroy_2013b}.

This indicates that parameters of the two-box model estimated from a shorter period could overestimate $\gamma$ and/or underestimate the domain depth and that these parameters should not be interpreted as general properties of the system. The parameters obtained by such a fit are still adequate in the temporal ranges over which the fit is performed and can be used to emulate climate of similar timescales. By virtue of having more parameters than its continuous diffusion form, the two-box model is more adequate in reproducing the time scales over which it has been calibrated. One should be careful if for example the parameters are estimated over a 10-year period and the two-box model is integrated for 100 years. Another culprit of diffusion models is that the system response depends on the initial state and it cannot be initialized with $T_1=T_2=0$ when the system is at disequilibrium $T_1 \neq T_2$.

\subsection{Simple estimates of expected observational trends}

We test the analytical expressions from equations \ref{eq:qss_approach}-\ref{eq:qss_approachn} against the observed trends in global mean surface temperature and top of atmosphere imbalance. Taking the time derivatives of $T(t)$ and $N(t)$, one can estimate their tendencies under linearly increasing forcing $F(t) = at$ as:

\begin{equation}
\frac{\dd T}{\dd t} = \frac{a}{\lambda} - \frac{e^{-t/\tau}}{\tau} \frac{aC}{\lambda^2}
\end{equation}
\begin{equation}
\frac{\dd N}{\dd t} = \frac{e^{-t/\tau}}{\tau} \frac{aC}{\lambda}
\end{equation}

\citet{Myhre_2025} do point out that within climate models $\frac{\dd N}{\dd t} \sim \frac{1}{\lambda}$.

The historical forcing from greenhouse gasses and aerosols  \citep{cmip6_forcing} can be approximated to a long period with weak linear forcing and a ramp up in the 1970s: from 1750 until 2025 (275 years) with a growth rate $a_1 = 0.001$~Wm$^{-2}$year$^{-1}$ and from 1970 to 2025 (55 years) with a growth rate $a_2 = 0.051$~Wm$^{-2}$year$^{-1}$. Since the model is linear, the estimate at year 2025 can be obtained as a sum of the estimates for the two growth rates $a_1$ and $a_2$ at the year 2025.

Supposing realistic values of $\lambda$, $C$ and $\tau$ can yield an order of magnitude estimate of the current state of the system and evaluate the structure of the diffusive model. Let $\lambda =1$Wm$^{-2}$K$^{-1}$, $C=10^{10}$Jm$^2$K$^{-1}$, $K=350$Wm$^{-1}$K$^{-1}$, yielding $\tau\approx 1000$ years, then for the year 2025 one could expect a total temperature change of $2.1$K, a temperature tendency of $0.36$K/decade, a TOA imbalance of $1.0$~Wm$^{-2}$ and a trend in the TOA imbalance of $0.17$~Wm $^{-2}$/decade. For comparison, the observed values which might be influenced by decadal-scale internal climate variability are $T\approx 1.4$K, $\frac{dT}{dt}\approx 0.28$K/decade (based on the the ERA5 global reanalysis \citep{Hersbach_2020} but the trends vary across datasets \citep{Menemenlis_2025}), a TOA imbalance of $N=1.6 \pm 0.2$~Wm$^{-2}$, $dN/dt = 0.45 \pm 0.18$~Wm $^{-2}$/decade \citep{Loeb_2024}. 

To an order of magnitude, these estimates agree with observations. Adjusting the two parameters in the energy balance framework discussed here: $\lambda$ and $K$, cannot provide a better estimate of $T$, $N$ and their tendencies. A better agreement can be obtained if $\lambda$ is not a constant parameter but a time-dependent unknown function: this demonstrates the puzzling fact that the effective climate feedbacks have been stronger prior to the 1960s and have since weakened (e.g. \citet{Andrews_2018, Fueglistaler_2021, Lewis_2021, Salvi_2023, Cael_2023}). Supposing that a bias in the observations or forcing estimates is unlikely and the influence of decadal-scale variability is not of such a magnitude, one is left to reshape the theoretical framework. Reshaping can include an  efficacy factor (as in \citet{Winton_2010, Held_2010}), which results in a time-evolving effective climate feedback modulated by the strength of ocean heat uptake. Another meaningful modification would be a more complex representation of vertical ocean mixing with an additional mixing pathway independent of the forcing-induced temperature difference as recently proposed by \citet{Gregory_2023}. Having a tropical and high-latitude pathways is motivated by the fact that ocean heat uptake in the high latitudes (e.g. AMOC) does not depend on global mean surface temperature that is governed mainly by tropical temperatures \citep{Gregory_2023}. Or one can utilize a different geometric structure that links the ocean's vertical and meridional stratification as in \citet{Marshall_2014}, again allowing the deeper layers to warm not only by the heat flux from the upper layers. A model with both vertical and meridional heat exchange can be made to represent a time-dependent effective climate feedback with a mean temperature component and a temperature stratification component as identified in observations \citep{Fueglistaler_2019}.

\section{Summary}

Based on energy balance models which describe the atmosphere-ocean system using a surface temperature feedback and vertical heat exchange below, we use the heat (diffusion) equation with energy-conserving boundary conditions to provide a simple analytical description of the transient climate response. The two parameters which define the system response are $\lambda$ and $K$. The surface temperature feedback $\lambda$ is a property of the atmosphere and the heat conductivity coefficient $K$ represents the ocean mixing. The heat conductivity in depth should be interpreted as a dynamical response of the fluid to a potential difference (here, temperature) rather than diffusion. The solution to this equation in response to linearly increasing forcing $\frac{dF}{dt} = a$ can be approximated to:

$$
\begin{aligned}
 T(t) &\approx \frac{a}{\lambda}t - (1-e^{-t/\tau})\frac{acL}{\lambda^2} \\
 N(t) &\approx (1-e^{-t/\tau})\frac{acL}{\lambda}
\end{aligned}
$$

with a time scale $\tau \approx \frac{c}{K}\frac{4L^2}{\pi^2}$, which depends on the mixing coefficient $K$ and the domain depth squared $L^2$ and is on the order of thousands of years. 

We can relate this solution to the conventional definition of $T_{\mathrm{TCR}}$ as the temperature at the time of doubling of forcing in a 1pctCO2 scenario in AOGCM simulations. The response time scale $\tau$ depends on $K$ and $L$ and it will vary between AOGCMs. In order to derive parameters that describe the intrinsic system properties from the conventional definition of $T_{\mathrm{TCR}}$, one needs to account for the AOGCM-dependent timescale of response. The 150-year CMIP6 simulations are too short to estimate the total heat capacity of the system and we assume $L=2600$m based on the effective ocean depth. This allows us to estimate the heat exchange coefficient $K$ from the 1pctCO2 simulations given $F_{2x}$ and $\lambda$ ($F_{2x}$ and $\lambda$ can be calculated from the abrupt-4xCO2 scenario). To derive $K$, we need to normalize the linear trend (derivative) of the surface temperature evolution in a 1pctCO2 simulation to the derivative of the corresponding steady state temperature $\frac{\dd T_\mathrm{SS}}{\dd t}=\frac{a}{\lambda}$, which removes the dependence on the forcing rate $a$. In a short period immediately after the forcing onset, over which the evolution of the surface temperature is quasi-linear, this is equivalent to taking $T_{TCR}/T_{ECS}$. The suppression of the surface warming rate is time-dependent and depends on $K$ through the system response time scale, and on $\lambda$ through the offset between the steady state ($T_\mathrm{SS}$) and quasi-steady state temperatures ($T_\mathrm{QSS}$):

\begin{align*}
\frac{\dd T}{\dd t}(t) \Big/ \frac{\dd T_\mathrm{SS}}{\dd t} = \begin{cases} 
   1 - \frac{K \pi^2}{4 \lambda L}, ~t\rightarrow0\\
   1, ~t/\tau >> 1.
\end{cases}
\end{align*}

It evolves from around 0.67 (AOGCM-dependent) near the forcing onset to 1 in quasi-steady state (equation \ref{eq:for_k} and Table \ref{tab}). In other words, before reaching quasi-steady state over thousands of years, the surface temperature would evolve further and further away from $T_\mathrm{SS}$ and the TOA imbalance will increase even for forcing with a linear growth. 

Evaluating this time-dependent ratio $\frac{\dd T}{\dd t}(t) \big/ \frac{\dd T_\mathrm{SS}}{\dd t}$ at a fixed time after the forcing onset $t=t'$ allows us to estimate the ocean mixing coefficient $K$ to be on the order of $350$~Wm$^{-1}$K$^{-1}$ in climate models. This would correspond to two-box model $\gamma \approx 0.4$~Wm$^{-2}$K$^{-1}$ or ocean diffusivity $\kappa = \frac{K}{c} \approx 2 \times 10^5$ m$^2$s$^{-1}$. AOGCMs with a smaller $K$ have a higher $\frac{\dd T}{\dd t}(t) \big/ \frac{\dd T_\mathrm{SS}}{\dd t}$ and $T_{\mathrm{TCR}}$. This estimate of $K$ might be biased high because in a real ocean $K=K(z)$ and a 150-year simulation allows us to only probe the first $\sim 500$ m of the ocean.

The two-box model \citep{Gregory_2000, Held_2010} can be derived from the form of the heat equation discussed here -- both are, in essence, a diffusion model forced at the top and with a surface temperature feedback.

Such an energy balance framework gives a good order of magnitude estimate for the observed surface temperature and TOA imbalance tendencies. The framework is structurally limited because it does not include time-varying effective climate feedback (as is done by the inclusion of an efficacy factor in \citet{Held_2010}). Another structural improvement could be including an ocean heat uptake pathway which does not depend on the temperature (potential) difference, such as the high-latitude pathway in \citet{Gregory_2023}. In an improved framework, coupling-specific parameters would be needed in addition to the atmosphere-specific $\lambda$ and ocean-specific $K$, with the effective climate feedback depending on all of these parameters.

%%%%%%%%%%%%%%%%%%%%%%%%%%%%%%%%%%%%%%%%%%%%%%%%%%%%%%%%%%%%%%%%%%%%%
% FIGURES---INSERT NEAR IN-TEXT DISCUSSION
%%%%%%%%%%%%%%%%%%%%%%%%%%%%%%%%%%%%%%%%%%%%%%%%%%%%%%%%%%%%%%%%%%%%%
%%  Enter figures near where they are discussed within the document.
%%
%
%\begin{figure}[t]
%  \noindent\includegraphics[width=19pc,angle=0]{figure01.pdf}\\
%  \caption{Enter the caption for your figure here.  Repeat as
%  necessary for each of your figures. Figure from \protect\cite{Knutti2008}.}\label{f1}
%\end{figure}

\clearpage
%%%%%%%%%%%%%%%%%%%%%%%%%%%%%%%%%%%%%%%%%%%%%%%%%%%%%%%%%%%%%%%%%%%%%
% ACKNOWLEDGMENTS
%%%%%%%%%%%%%%%%%%%%%%%%%%%%%%%%%%%%%%%%%%%%%%%%%%%%%%%%%%%%%%%%%%%%%
\acknowledgments

I thank Stephan Fueglistaler for constructive input and feedback that shaped this paper. I am also grateful to Doris Folini for her continuous enthusiasm for math problems. The numerical solution of the heat equation was realized on the GPU machine set up and maintained by Daniel Tsvetkov and Denitsa Rusinova as a personal project.

I am grateful for the thoughtful remarks of the three anonymous reviewers, which greatly improved the article.

The WCRP, CMIP6 modeling groups, and ESGF are acknowledged for enabling and providing the data.

The data analysis and visualization have been performed using free and open source software: a big thanks to the community for building powerful tools from which everyone can benefit.

B.C. is supported by Swiss National Science Foundation Grant P500PN\_222086.

%%%%%%%%%%%%%%%%%%%%%%%%%%%%%%%%%%%%%%%%%%%%%%%%%%%%%%%%%%%%%%%%%%%%%
% DATA AVAILABILITY STATEMENT
%%%%%%%%%%%%%%%%%%%%%%%%%%%%%%%%%%%%%%%%%%%%%%%%%%%%%%%%%%%%%%%%%%%%%
% 
%
\datastatement

Wolfram Mathematica notebooks and analysis code are available at \url{https://gitlab.com/mjade/analytical-insights-into-tcr}. The version used in this study corresponds to commit bbb7ac7e.

The CMIP6 archive can be accessed at \url{https://esgf-node.llnl.gov/projects/cmip6/}.

\appendix[A] 
\appendixtitle{ Derivation of slowest decaying mode}

% p. 11 of notes has the derivation of the SS solution
Equation \ref{eq:qss} is the quasi-steady state solution of the heat equation (equation \ref{eq:diffusion}) with boundary conditions from equation \ref{eq:bcs}, i.e. this is the long-term limit of the solution. We also want to obtain the full time-dependent solution to the heat equation, which describes the departure from the initial condition. For an unperturbed state in equilibrium (steady state), the initial condition would be a constant profile in depth, which we set to 0:
\begin{equation}
\mathcal{T}(t=0, z) = 0    
\end{equation}

In this case the full solution is given by:
\begin{equation}
\label{eq:full}
    \mathcal{T}(t, z) = -\frac{a}{\lambda}t + \sum_{k_1=1}^{\infty} f(k_2(k_1)) e^{-\frac{K}{C} k_2(k_1)t},
\end{equation}
where the coefficients $k_2$ are all possible roots of the transcendental equation:
\begin{equation}
\label{eq:transcendental}
     tan(L\sqrt{k_2}) = \frac{\lambda}{K\sqrt{k_2}}
\end{equation}

The first root of equation \ref{eq:transcendental} is the largest and will give us the slowest decay mode $e^{-\frac{t}{\tau_1}}$, where the decay time is $\tau_1=\frac{c}{K k_2(1)}$. We can solve for $k_2(1)$ by finding the first crossing point for the curves $tan(L\sqrt{x})$ and $\frac{\lambda}{K\sqrt{x}}$. Since all parameters are positive, we let $u=\sqrt{x}/L$ and then the curves become $f_1(u) = tan(u)$ and $f_2(u) = L\lambda/(Ku)$. In the limit $L \lambda>>K$, the first root approaches $\pi/2$. In the climate case, the climate feedback parameter $\lambda$ is around 1Wm$^{-2}$K$^{-1}$; the depth of the ocean $L$ is on the order of a few thousand of meters and the effective ocean diffusivity $K$ is usually a few hundred Wm$^{-1}$K$^{-1}$. This gives us the slowest decaying mode: $\tau_{sl} = \tau_1 \approx \frac{c}{K}\frac{4L^2}{\pi^2}$.

If one further assumes that this is the only time-dependent mode which acts on the system to approach quasi-steady state, the approximation for the full solution can be written as equation \ref{eq:qss_approach}:
$$
    \mathcal{T}(t, z) \approx \frac{a}{\lambda}t - (1-e^{-t/\tau_{sl}})f(z),
$$

where $f(z) = \frac{-2 a c K L - 2 a \lambda c L z + a \lambda c z^2}{2 \lambda^2 K}$ is the quadratic depth profile that needs to be developed towards quasi-steady state.

This approximation can become invalid for the initial departure ($t \rightarrow 0$) from the unperturbed profile in the deeper layers ($z \rightarrow L$) for a large forcing rate $a$. If $a$ is larger than the initial phase given by $(1-e^{-t/\tau_1})f(z)$, then the temperature at levels where the QSS $f(z)$ is small (i.e. $z \rightarrow L$) might become negative. This solution is unrealistic and is due to the fact that we have neglected all other depth-dependent decaying time scales $\tau_2, \tau_2,..$ that correspond to $i=2, 3...$ in the full solution (equation \ref{eq:full}).

This approximation can also underestimate $\tau$ if $K$ is sufficiently large, and the crossing point of $f_1(u)$ and $f_2(u)$ is before $\frac{\pi}{2}$. 

Figure \ref{fig:ts} shows both the full solution (evaluated numerically) and the approximation (evaluated analytically). The solution for the upper layers ($z \rightarrow 0$) given by equation \ref{eq:qss_approach} is sufficiently accurate and the non-perfect match between the simulated temperature profile in depth (numerical solution) and its analytical solution at QSS (Figure \ref{fig:ts} b) is due to a slight underestimation of $\tau$ as mentioned above.

In comparison to the exact diffusion solution, the slowest mode approximation will overestimate the temperature tendency (Figure \ref{fig:ts} c and Figure \ref{fig:cm4} a) and underestimate the imbalance tendency (Figure \ref{fig:imb} b and Figure \ref{fig:cm4} b). 

The slowest mode approximation is only appropriate for the 1pctCO2 case in which the temperature evolution has two time-evolving components: $\frac{a}{\lambda}t$, and a relaxation to QSS: $(1-e^{-t/\tau_{sl}})f(z)$, the first one being dominant. For an abrupt-4xCO2 case, the system relaxation cannot be adequately described with a single time scale: the exact solution of the continuous diffusion equation with a surface temperature feedback possesses a continuum of time scales which project onto the surface temperature as many fast and slow components, similar to the fast and slow component in the two-box model \citep{Gregory_2000, Held_2010}.

\newpage
\appendix[B]
\appendixtitle{Analogy to the two-layer model}

\begin{figure}[h!]
    \centering
    \includegraphics[width=\linewidth]{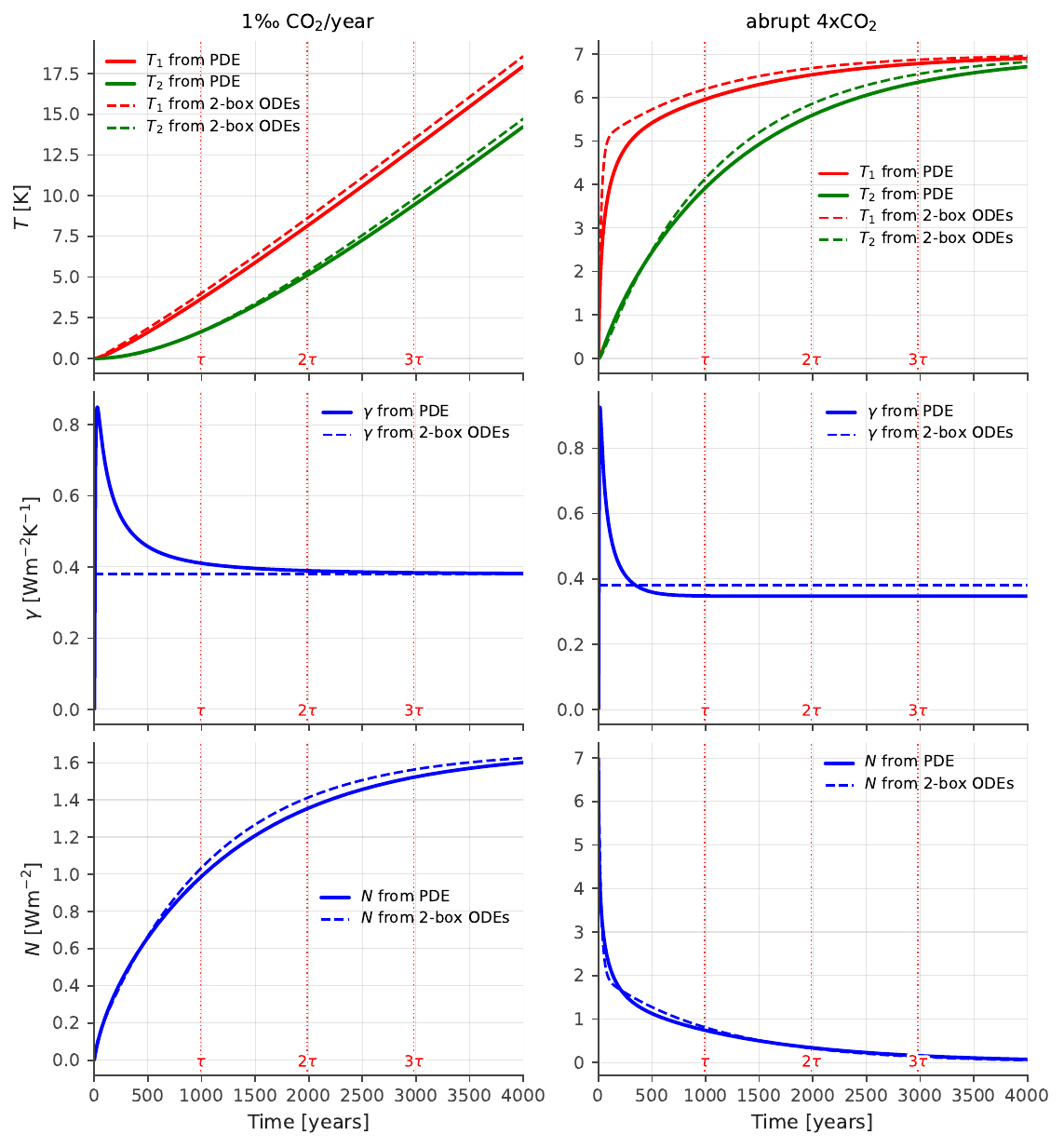}
    \caption{Numerical solutions of the heat equation and the 2-layer model. Parameters for the heat equation are the same as in Figure \ref{fig:ts}; equivalent parameters for the 2-box model are chosen for $k=0.1$ and $L=2600$ m as $C_1=260$ m, $C_2=2340$ m, the equivalent $\gamma$ is estimated to be $0.38$ Wm$^{-2}$K$^{-1}$.}
    \label{fig:2box}
\end{figure}

We use the QSS expression (equation \ref{eq:qss}) to derive an equivalent expression for the two-box model (equation \ref{eq:two-box}), and we evaluate the two-box model's approach to QSS numerically.

We will use the continuous diffusion partial differential equation (equations \ref{eq:diffusion}, \ref{eq:bcs}) to derive the two-box model of two coupled ordinary differential equations (equation \ref{eq:two-box}). The temperatures of the two layers in the two-box model correspond to the vertically averaged temperatures in the continuous diffusion model, split at a certain depth ratio:
\begin{align}
    \label{eq:integrals_2box_T1}
    T_1 &= \int_0^{kL} \mathcal{T}(t, z) dz / (kL-0)\\
    \label{eq:integrals_2box_T2}
    T_2 &= \int_{kL}^{L} \mathcal{T}(t, z) dz / (L-kL)
\end{align}

The QSS solution to equations \ref{eq:integrals_2box_T1} and \ref{eq:integrals_2box_T2} is:
\begin{align}
    T_1 &= \frac{a}{\lambda}t + \frac{acL(-6K+\lambda(-3+k)kL)}{6 \lambda^2 K}\\
    T_2 &= \frac{a}{\lambda}t + \frac{acL(-6K+\lambda(-2+(-2+k)k)L)}{6 \lambda^2 K}
\end{align}

Obtaining $T_1$ and $T_2$ from the QSS solution $T(t, z)$ in equation \ref{eq:qss} allows us to backwards engineer $\gamma$ from the original system (equation \ref{eq:two-box}) as:

\begin{align}
    C_2 &= (L - kL) c \\
    \label{eq:gamma}
    \gamma_\mathrm{QSS} &= \frac{C_2 \frac{\partial T_2}{\partial z}}{T_1 - T_2} = K \frac{6 (k-1)}{L (k-2)}
\end{align}

This equation links $\gamma$ [Wm$^{-2}$K$^{-1}$] to the heat conductivity $K$ [Wm$^{-1}$K$^{-1}$] and the coefficient $k$ which splits the continuous finite domain of the heat equation solution to two finite layers. Based on fits to climate models in \citet{Geoffroy_2013}, where $L \approx 800$ m and the depth of the upper layer is $\sim 80$ m, then $k \approx 0.1$. 

The equation is time independent because it is based on the steady state solution, which allows us to obtain a clear expression. In the approach to steady state $\gamma$ will evolve over time and approach the predicted value.

As seen in Figure \ref{fig:2box} when the heat equation is reduced to the 2-box model, $\gamma$ is larger in the beginning and approaches $\gamma_\mathrm{QSS}$ from equation \ref{eq:gamma}, which we estimate analytically. Since $\gamma$ is usually estimated from coupled models, i.e. done the other way around, it is likely that the fitted parameter is biased high if one wants to compare it to the heat equation.

We briefly note that the heat capacities have different units: in equation \ref{eq:diffusion} we use the volumetric heat capacity $c$ [Jm$^{-3}$K$^{-1}$], while in the two-box model (equation \ref{eq:two-box}) the heat capacities relate the temperature tendencies directly to energy fluxes, and hence have the units  $C_1, C_2$ [Jm$^{-2}$K$^{-1}$], which is sometimes expressed as $C_1, C_2$ [W year$^{-1}$ m$^{-2}$ K$^{-1}$] \citep{Geoffroy_2013}.

\newpage
\appendix[C]
\appendixtitle{The apparent heat capacity for $F=at$ and $F=const$}

The apparent heat capacity is a useful mathematical abstraction which relates the system imbalance to the surface temperature tendency. The temporal evolution, lower and upper limits of the apparent heat capacity depend on the type of diffusion model (two-box or continuous) and on the functional form of the forcing. We derive these limits for the case of linearly increasing forcing $F=at$ (e.g. 1pctCO2) and the Heaviside case of abrupt constant forcing $F=const$ (e.g. abrupt-4xCO2).

Starting from the two-box model as in equation \ref{eq:two-box}, one can diagnose the apparent heat capacity from the total temperature tendency of the system:

\begin{equation}
    N = C_1 \frac{dT_1}{dt} + C_2 \frac{dT_2}{dt}
\end{equation}

\begin{equation}
    C_\mathrm{app} = \frac{N}{dT_1/dt} = C_1 + C_2 \frac{dT_2}{dT_1}
\end{equation}

And normalize to the volumetric heat capacity of seawater $c=4 \times 10^6$ [Jm$^{-3}$K$^{-1}$] to get the apparent ocean depth $L_\mathrm{app} = C_\mathrm{app}/c$ in meters.

For $F=at$ the upper limit of $C_\mathrm{app}$ is QSS, where $\frac{dT_1}{dt}=\frac{dT_1}{dt}=\frac{a}{\lambda}$, therefore $C_\mathrm{app} = C_1 + C_2$, which is the domain depth or total heat capacity of the system.

For $F=const$, which is a Heaviside relaxation case, as time evolves the temperature tendency in the deeper layer $\frac{dT_2}{dt}$ becomes larger than the one of the upper layer $\frac{dT_1}{dt}$ and they both decrease with time as the system approaches flux equilibrium. An interesting feature is that the ratio of the temperature tendencies is constant and therefore $C_\mathrm{app}$ has an upper limit that is reached at some point in time before flux equilibrium. The upper limit of the ratio of the temperature tendencies is:

\begin{equation}
\label{eq:dt2dt1}
    \frac{dT_2}{dT_1} = \frac{-\gamma C_{1} + \sqrt{(\gamma C_{1} + C_{2}(\gamma \epsilon + \lambda))^{2} - 4\gamma C_{1}C_{2}\lambda} + \gamma C_{2}\epsilon + C_{2}\lambda}{2\gamma C_{2}\epsilon},
\end{equation}

which for $C_{1} \rightarrow 0$ evaluates to $\lim_{\scriptscriptstyle C_1 \to 0} \frac{dT_2}{dT_1} = \frac{\lambda + \epsilon \gamma}{\epsilon \gamma}$.

Since $\frac{dT_2}{dt} > \frac{dT_1}{dt}$, the upper limit of the apparent heat capacity for the Heaviside ($F=const$) forcing form will be higher than the total system heat capacity.

\begin{figure}[h!]
    \centering
    \includegraphics[width=\linewidth]{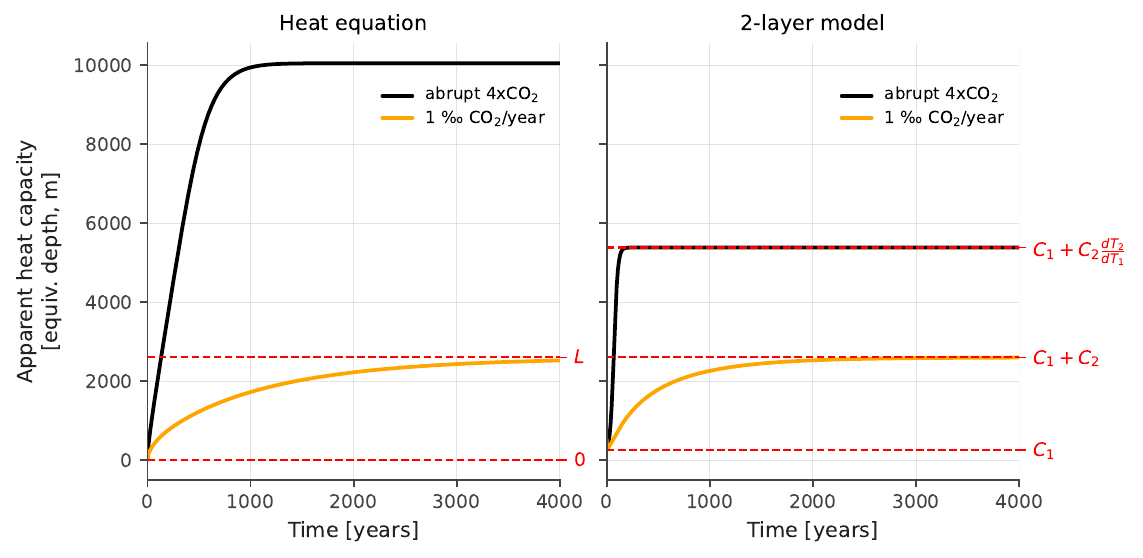}
    \caption{The apparent heat capacity as equivalent depth for different functional forms of the forcing diagnosed from the numerical solution of the heat equation (left) and diagnosed from the two-layer model (right). Model parameters are the same as in Figure \ref{fig:2box}. $dT_2/dT_1$ is calculated from equation \ref{eq:dt2dt1} with $\epsilon=1$.}
    \label{fig:capp}
\end{figure}

The temporal evolution of $C_\mathrm{app}$ for the continuous diffusion PDE and 2-layer model with both forms of forcing is shown in Figure \ref{fig:capp}.

The lower limit of $C_\mathrm{app}$ is 0 for the continuous diffusion model (heat equation) and $C_1$ for the two-layer model.

Within AOGCMs, the limits of $C_{\mathrm{app}}$ cannot be readily diagnosed. The lower limit needs to be done over a very short period close to the beginning of the simulation, in which the role of internal variability would be substantial. Figure \ref{fig:cm4} c shows the apparent heat capacity as diagnosed from the GFDL-CM4 1pctCO2 simulation, which points to a continuous evolution of $C_{\mathrm{app}}$ with a lower limit below 13 W yr m$^{-2}$K$^{-1}$ diagnosed around year 15 ($C_{\mathrm{app}}=13$W yr m$^{-2}$K$^{-1}$ corresponds to an effective ocean depth $L_{\mathrm{app}}\approx 100$m).

%%%%%%%%%%%%%%%%%%%%%%%%%%%%%%%%%%%%%%%%%%%%%%%%%%%%%%%%%%%%%%%%%%%%%
% REFERENCES
%%%%%%%%%%%%%%%%%%%%%%%%%%%%%%%%%%%%%%%%%%%%%%%%%%%%%%%%%%%%%%%%%%%%%

\bibliographystyle{ametsocV6}
\bibliography{references}

\end{document}